\documentclass{nature}

\usepackage{amsmath,bm}
\usepackage[version=3]{mhchem}
\usepackage[normalem]{ulem}

\title{Activating spin-forbidden transitions in molecules by the highly
localized plasmonic field}

\author{Sai Duan$^{1,2}$, Zilvinas Rinkevicius$^{2,3}$ and Yi
Luo$^{1,2\star}$}

\begin{document}

\maketitle

\begin{affiliations}
\item Hefei National Laboratory for Physical Sciences at the Microscale,
Synergetic Innovation Center of Quantum Information \& Quantum Physics, 
University of Science and Technology of China, Hefei, 230026
Anhui, P. R. China.
\item Division of Theoretical Chemistry and Biology, School of
Biotechnology, KTH Royal Institute of Technology, S-106 91 Stockholm,
Sweden.
\item Swedish e-Science Research Centre, KTH Royal Institute of
Technology, S-100 44 Stockholm, Sweden.
\end{affiliations}

\begin{abstract}
Optical spectroscopy has been the primary tool to study the electronic
structure of molecules\cite{book1,book2}. However the strict spin
selection rule has severely limited its ability to access states of
different spin multiplicities. Here we propose a new strategy to
activate spin-forbidden transitions in molecules by introducing
spatially highly inhomogeneous plasmonic field. The giant enhancement of
the magnetic field strength resulted from the curl of the inhomogeneous
vector potential makes the transition between states of different spin
multiplicities naturally feasible. The dramatic effect of the
inhomogeneity of the plasmonic field on the spin and symmetry selection
rules is well illustrated by first principles calculations of \ce{C60}.
Remarkably, the intensity of singlet-triplet transitions can even be
stronger than that of singlet-singlet transitions when the plasmon
spatial distribution is comparable with the molecular size. This
approach offers a powerful means to completely map out all excited
states of molecules and to actively control their photochemical
processes. The same concept can also be applied to study nano and
biological systems.
\end{abstract}

Optical excitation is the fundamental process that controls molecular
properties and their spectroscopies. In spite of great success, the
optical excitation with conventional light sources could only access to
very limited number of excited states due to the restriction of the
intrinsic symmetry and spin selection rules. It was reported recently
that the symmetry selection rule for optical excitations could be
softened under the plasmonic
field\cite{andersen2011np,filter2012prb,jain2012pnac,nobusada2013,
takase2013np,trinh2013nl,yannopapas2015jmo,yang2016pra,rivera2016science},
which takes advantage of its near field characteristic. We will
demonstrate here that the use of locality of the plasmonic field could
not only further weaken the symmetry selection rule, but also remove
completely the much stricter spin selection rule. 

Within non-relativistic regime, the light-matter interaction is governed
by the minimal coupling
Hamiltonian\cite{dirac1927prsla1,dirac1927prsla2}, which is adequate
even for plasmonic fields\cite{rivera2016science,
altewischer20002nature,archambault2010prb,tame2013nphy,tielrooij2015np}.
In this context, the light-matter interaction Hamiltonian,
$\hat{\mathcal{H}}^\prime$, can be expressed
as\cite{dirac1927prsla1,rivera2016science} the summation of the vector
potential component $\hat{\mathcal{H}}_{\text{A}}^\prime$ and magnetic
component $\hat{\mathcal{H}}_{\text{B}}^\prime$
\begin{equation}
\begin{split}
\hat{\mathcal{H}}^\prime=&\hat{\mathcal{H}}_{\text{A}}^\prime+\hat{\mathcal{H}}_{\text{B}}^\prime\\
\hat{\mathcal{H}}_{\text{A}}^\prime=&\sum_k\frac{1}{2}\left[
\hat{\mathbf{p}}_k\cdot\mathbf{A}(\mathbf{r}_k,t)
+\mathbf{A}(\mathbf{r}_k,t)\cdot\hat{\mathbf{p}}_k\right]\\
\hat{\mathcal{H}}_{\text{B}}^\prime=&\sum_k\frac{1}{2}\hat{\bm{\sigma}}_k\cdot\mathbf{B}(\mathbf{r}_k,t)
\end{split}
\label{eq:Hint}
\end{equation}
where $\hat{\mathbf{p}}_k$ is the momentum operator of the $k$th
electron, $\mathbf{A}(\mathbf{r}_k,t)$ is the vector potential of the
light, $\mathbf{r}_k$ is the position of the $k$th electron,
$\hat{\bm{\sigma}}_k$ is the Pauli matrix for the $k$th electron, and
$\mathbf{B}(\mathbf{r}_k,t)$ is the magnetic field that can be
calculated by the curl of the vector potential\cite{jackson1998},
\textit{i.e.},
$\mathbf{B}(\mathbf{r}_k,t)=\nabla_k\times\mathbf{A}(\mathbf{r}_k,t)$.
Here we neglect the high-order $\mathbf{A}^2(\mathbf{r}_k,t)$
term\cite{dirac1927prsla2,filter2012prb}. Thus, the light-matter
interaction can be determined by the vector potential. In general,
vector potential could be expressed as (Figure~1a)
\begin{equation}
\mathbf{A}(\mathbf{r},t)=A_0g(\mathbf{r})
e^{\imath\eta_0\mathbf{k}_0\cdot\mathbf{r}}
e^{-\imath\omega{t}}\hat{\mathbf{n}}+\text{c.c.}
\label{eq:agen}
\end{equation}
where $A_0$ is the constant field amplitude, $g(\mathbf{r})$ is the
amplitude of the spatial distribution function, $\mathbf{k}_0$ is the
wave vector in vacuum, $\eta_0$ is the confinement factor for the wave
vector caused by surrounding dielectric media\cite{rivera2016science},
$\omega$ is the frequency of electromagnetic field, $\hat{\mathbf{n}}$
is the norm of the vector potential, and $\text{c.c.}$ is the complex
conjugate. As a result, we have
\begin{equation}
\mathbf{B}\left(\mathbf{r},t\right)=
\left[\imath\eta_0g(\mathbf{r})\mathbf{k}_0\times\hat{\mathbf{n}}
+\nabla{g(\mathbf{r})}\times\hat{\mathbf{n}}\right]
e^{\imath\eta_0\mathbf{k}_0\cdot\mathbf{r}}
e^{-\imath\omega{t}}+\text{c.c.}
\label{eq:bfield}
\end{equation}
According to the Fermi's golden rule\cite{zettili2001}, the absorption
cross section between molecular initial state $|i\rangle$ and final
state $|f\rangle$ caused by the interaction Hamiltonian reads
\begin{equation}
\sigma_{if}=2\pi\left|\langle{i}|\mathcal{H}^\prime|f\rangle\right|^2/\mathcal{F}
\label{eq:gif}
\end{equation}
where $\mathcal{F}$ is the flux of the incident light.

In general, matrix elements of light-matter interaction Hamiltonian in
Eq.~\ref{eq:gif} between arbitrary state of system are non-vanishing,
and the magnitude of these matrix elements depends on the nature of
vector potential and magnetic field of electromagnetic field entering
$\hat{\mathcal{H}}^\prime$. The vector potential component of
electromagnetic field governs the behaviour of the first
spin-independent term in $\hat{\mathcal{H}}^\prime$, and the magnetic
field component controls the second spin-dependent term, respectively.
This separation in $\hat{\mathcal{H}}^\prime$ leads to distinct
selection rules for its matrix elements, where the vector potential
component is responsible for selection rules for matrix elements between
states of the same multiplicity, while the magnetic field component
responsible for selection rules between states of different
multiplicities. For instance, for plane wave, $g(\mathbf{r})=1$, the
dipole approximation assumes
$e^{\imath\eta_0\mathbf{k}_0\cdot\mathbf{r}}\approx1$. With the help of
the quantum mechanical relationship
$\hat{\mathbf{p}}=\imath[\hat{\mathcal{H}}_0,\mathbf{r}]$, where
$\hat{\mathcal{H}}_0$ is the unperturbed molecular Hamiltonian, the
absorption between the same multiplicity is proportional to
$\left|\langle{i}|\sum_k\mathbf{r}_k|f\rangle\right|^2$, resulting in
the spatial symmetry selection rule (also called as the ``dipole
selection rule'')\cite{dirac1927prsla2,zettili2001}, \textit{i.e.}, only
the totally symmetric representation of the direct product of the
symmetry irreducible representation of $\mathbf{r}$, initial and final
states is allowed. Meanwhile, the involvement of the plan wave
($g(\mathbf{r})=1$) eliminates the gradient term for the magnetic field
(Eq.~\ref{eq:bfield}). Moreover, within the dipole approximation, the
magnetic field would become zero. As a result, the transitions between
different multiplicities are strictly forbidden, \textit{i.e.} the
so-called spin symmetry selection rule. Apparently, there are two ways
to break-down the spin symmetry selection rule. One is to go beyond the
dipole approximation, while another is to introduce a large confinement
factor that can lead to non-zero magnetic field. 

It is well established that with the conventional optical excitation
sources, none of these two approaches could be utilized. However, the
spatially confined plasmon (SCP) generated either by optical excitation
or electron current has offered exciting new opportunity. It has been
shown that the inhomogeneous plasmonic field can be confined in a
nano-cavity\cite{stockle2000cpl,zhang2013nature}. With such a highly
confined plasmonic field, the inner structure of a porphyrin molecule
has been visualised by the tip-enhanced Raman spectroscopy (TERS) with a
super high spatial resolution of 0.5~nm\cite{zhang2013nature}. Our
recent theoretical work has nicely reproduced the experimental Raman
images of the molecules by taking into account the spatial distribution
of the SCP in the transition matrix\cite{duan2015jacs}. It was found
both theoretically and experimentally that the spatial distribution of
the SCP has been confined within the size of 2~nm, which is comparable
with the size of many molecules\cite{duan2015jacs,jiang2015nn}. In
principle, the spatial size of the SCP is determined by the Thomas-Fermi
screening length\cite{atkin2013nature}, which can be down to, for
example, 1~\AA{} for silver and gold\cite{kreibig1995}. With such a
highly localized field, a huge magnetic field is expected to be
generated through the curl of its vector potential, as indicated by
Eq.~\ref{eq:bfield}. In other words, even with vanished spin-orbit
coupling, the spin symmetry selection rule can be completely removed.
For the sake of the presentation, we name afterwards this new transition
as plasmon induced spin transition (PIST)

We will verify the actual magnitude of the effects from the PIST under
different confined plasmonic field by calculating the absorption
spectrum of a highly symmetric buckminsterfullerene \ce{C60}. The
computational methods are summarized in the Methods section. Due to its
very high symmetry $\text{I}_{\text{h}}$, under the dipole
approximation, only the ${}^1\text{T}_{\text{1u}}$ excited states of
\ce{C60} are allowed optical transitions\cite{cotton1990}, which give
rise to only one main absorption band (the C band) above 300~nm in the
experimental spectrum\cite{coheur1996jpbamop}. Although the inclusion of
the Herzberg-Teller (for singlet) and spin-orbit coupling (for triplet)
could break the selection rules, their effects only make very small
contribution to the low energy transitions as illustrated in previous
experiments (black line in Figure~1b)\cite{coheur1996jpbamop}. Without
considering both vibronic and spin orbital couplings, our calculated
spectrum does well reproduce the experimental one as shown in Figure~1b.
The orientation of the molecule is an important fact when it interacts
with the highly localized plasmonic field. For the sake of the
simplicity, we have adopted the 6-Ring configuration of \ce{C60} on a
spaced substrate (Figure~1a) as mostly observed in previous
experiments\cite{li2009prl,berland2013prb}. The C band is again dominant
above 300~nm in the absorption spectrum of \ce{C60} on surface with the
6-Ring configuration with conventional light source shown in Figure~1c.
It should be noted that the absorption cross section from single
molecules can be experimentally measured through the photoluminescence
excitation technique\cite{zhang2013nature}.

The first consideration is the effects of the going beyond the dipole
approximation in wavevector expansion. The SCP is squeezed within a
cavity of nanometer size, which is equivalent to the shrinking of the
wavelength\cite{rivera2016science}. The calculated cross sections as
shown in Figure~2a clearly indicate that the shrinking of the wavelength
can significantly break the spatial symmetry selection rule, which is
consistent with what was proposed by Rivera \textit{et
al.}\cite{rivera2016science} It comes from the multipole contributions
in the expansion of $e^{\imath\eta_0\mathbf{k}\cdot\mathbf{r}}$
(Figure~2b). For instance, two $^1\text{T}_{\text{1g}}$ states from the
electric quadrupole contribution emerge when the wavelength is only
shrunk by a factor of 50. When $\eta_0=100$, the absorptions of
$^1\text{T}_{\text{1g}}$ bands are even larger than the dipole allowed C
band transition. It is noted that, with $\eta_0$ less than 100, the
breakdown of the spin selection rule shows a very weak sign, resulting
in negligible PIST. Again, these observations are consistent with the
previous theoretical prediction\cite{rivera2016science}. It is
interesting to note that the further shrinking of the wavelength by a
factor of 300 can generate strong PIST of the triplet
$^3\text{T}_{\text{1u}}$ state caused by the magnetic dipole, which
drastically breaks down the spin selection rule. It should be noted
that, even when $\eta_0$ equals 500, the dipole-forbidden singlet
$^1\text{T}_{\text{1g}}$ state around 600~nm is still the most intense
band, although more PISTs, for instance the magnetic quadrupole allowed
$^3\text{H}_{\text{g}}$, emerge.

The highly localized plasmonic field can result in a huge gradient of
the amplitude distribution of vector potential that is strongly depended
on the spatial distribution of the field, leading to the generation of a
much enhanced magnetic field. Our calculated absorption cross sections
of \ce{C60} excited by the plasmon of different spatial size without
shrinking the wavelength are depicted in Figure~3a. Here the plasmonic
size ($\Gamma$) is determined by the full width at half-maximum of field
distribution (see the Methods section). It is found that for the
plasmonic size of 20~\AA{}, the absorption spectrum is almost identical
to that from the plan wave excitation except a very weak signal from the
$^3\text{T}_{\text{1u}}$ state around 427~nm. This is reasonable since
the size of the plasmonic field is about 3 times of the size of \ce{C60}
around 7~\AA{}\cite{hedberg1991science}. When the plasmonic size
approaches the molecular size, for example 10~\AA{}, the PISTs of
$^3\text{T}_{\text{1u}}$ and $^3\text{H}_{\text{g}}$ states become
significant, nicely illustrating the breakdown of the spin symmetry
selection rule. Meanwhile, the appearance of the singlet transitions to
the $^1\text{H}_{\text{g}}$ states also indicates that the symmetry
selection rule is broken as well. When the molecule interacts with a
highly localized plasmonic field, the symmetry of the whole interactive
system is significantly reduced in comparison with the molecule itself,
which results in the breakdown of the symmetry selection rule. As shown
in Figure~3c, the symmetry of the whole interactive system reduces from
$\text{I}_{\text{h}}$ to $\text{C}_{\text{3v}}$. When the plasmonic size
sets to be the same of the \ce{C60} size (7~\AA{}), the
$^3\text{T}_{\text{1u}}$ state around 427~nm gains huge intensity and
becomes the most intense absorption band. Under this condition, the much
enhanced magnetic field induced by the gradient term in
Eq.~\ref{eq:bfield} kicks in (Figure~3b). Due to the same reason, all
singlet absorption transitions are suppressed by the intense PISTs when
the plasmonic size is set to 5~\AA{}. It should be stressed that the
small shrink of wavelength, for example setting $\eta_0=50$, only gives
minor affect for the absorption cross sections when the inhomogeneity is
also taken into account. Further shrinking ($\eta_0=100$) could indeed
enhance the singlet transitions but does not much affect PISTs
(Figure~4a and Figure~S1).

Noteworthy, when different positions of plasmon are adopted, the
symmetry of the whole system could be further reduced to even
$\text{C}_{\text{1}}$, where all transitions are allowed in principle.
Thus, the position-dependent absorption is expected. We depicted the
absorption cross sections under a 7~\AA{} plasmon without shrinking the
wavelength at different positions in Figure~4b for the most three
intense transitions mainly contributed by the transitions from highest
occupied molecular orbital (HOMO) to lowest unoccupied molecular orbital
(LUMO) of \ce{C60}. It should be noted that, all three states are near
degenerate (Table~S2, the maximum energy difference is 0.24~eV, which is
less than the experimental resolution of the C band, \textit{i.e.},
0.3~eV\cite{coheur1996jpbamop}, Figure~1b). It is nice to observe the
complementary patterns between different states. Specifically, the
${}^3\text{H}_{\text{g}}$ and ${}^1\text{H}_{\text{g}}$ have a bright
triangle and quasi-circular pattern located at the center six-member
ring, respectively. On the other hand, the central pattern of the
${}^1\text{T}_{\text{1g}}$ state is dark. The bright patterns separated
in real space (Figure~4b and Figure~S2) reveal that different states
could be selectively excited by precisely controlling of the plasmonic
position.

In summary, we theoretically demonstrate an efficient spin breaking
router by inhomogeneous plasmonic fields from the minimal coupling
Hamiltonian in non-relativistic regime. Taking the transitions for
singlet and triplet excited states of \ce{C60} under plasmonic fields as
an example, we find that the breakdown of the spin symmetry rule is
highly dependent on the size of the plasmons. When the size of plasmon
approaches to the molecular size, the transitions to triplet excited
states could be largely increased because of the enhanced magnetic field
contributed from plasmonic inhomogeneity. As a result, the absorption
cross sections of triplet transitions become comparable to or even
larger than that of the conventional dipole allowed singlet transitions.
In addition, the plasmonic position dependence of the absorption opens a
new pathway to manipulate different molecular quantum states in real
space. Our findings could be easily extended to Raman scattering, as
well as other linear and nonlinear optical processes, which could have
strong impact on different applications in chemistry, material science,
physics, and biology.

\begin{figure}
\caption{\textbf{Schematic illustration for \ce{C60} model system.}
\textbf{a}, Schematic figure for general vector potential of plasmonic
field confined in a nano-cavity formed between the tip and spaced
substrate. The insert shows the 6-Ring configuration of \ce{C60}
adsorbed on the substrate. \textbf{b}, Comparison between experimental
(black line) and theoretical (red line) absorption cross sections of
\ce{C60} in vacuum. The experimental data were obtained after
solid-vapour equilibrium under a stable temperature of 858~K by Coheur
\textit{et al.}\cite{coheur1996jpbamop} The theoretical spectrum was
convoluted by the Lorentzian function with a full width at the
half-maximum of 0.3~eV. The gray zone indicates the dipole and spin
forbidden region for absorption of \ce{C60}. \textbf{c}, Calculated
absorption cross sections of a single \ce{C60} adsorbed on a spaced
substrate with the 6-Ring configuration under the dipole approximation,
\textit{i.e.}, set the confinement factor ($\mathsf{\eta_0}$) to zero,
meanwhile, the full width at half-maximum of the amplitude distribution
function ($\Gamma$) to infinity. The red and blue bars on the top of
x-axis indicate the triplet and single transition energies,
respectively. The symmetry irreducible representations of all triplet
and singlet states mainly contributed by the transitions from highest
occupied molecular orbital to lowest unoccupied molecular orbital of
\ce{C60} are labeled in red and blue fronts, respectively. All
calculated cross sections were convoluted by the Lorentzian function
with a full width at half-maximum of 0.05~eV. All calculated transition
energies in \textbf{b} and \textbf{c} were shifted by 0.35~eV to compare
the experimental observations.}
\end{figure}

\begin{figure}
\caption{\textbf{Absorption under the plane wave plasmonic field.}
\textbf{a}, Calculated absorption cross sections of a single \ce{C60}
adsorbed on a spaced substrate with the 6-Ring configuration under the
plane wave plasmonic field with different confinement factor $\eta_0$
from 50 to 500 (bottom to top). The red and blue areas in absorption
cross sections represent the contributions from triplet and single
excited states, respectively. The symmetry irreducible representations
of all significant triplet and singlet transitions are labeled in red
and blue fronts, respectively. All calculated transition energies were
shifted by 0.35~eV and cross sections were convoluted by the Lorentzian
function with the full width at half-maximum of 0.05~eV. \textbf{b},
Schematic illustration of analytical and Taylor expansion up to
different orders for the mode function in the plane wave plasmonic
field. The real and imaginary parts are represented by the solid and
dotted lines, respectively. In Taylor expansion, corresponding symmetry
irreducible representations of allowed final states under different
orders are labeled.}
\end{figure}

\begin{figure}
\caption{\textbf{Absorption under the localized plasmonic field.}
\textbf{a}, Calculated absorption cross sections of a single \ce{C60}
adsorbed on a spaced substrate with the 6-Ring configuration under the
localized plasmonic field without shrinking wavelength at different
plasmonic size from 20 to 5~\AA{} (bottom to top). The plasmons are
placed on around 2~\AA{} above the center of the center six-member ring
and the size of the plasmon is determined by $\Gamma$ (the full width at
half-maximum of the amplitude distribution function). The red and blue
areas in absorption spectra represent the contributions from triplet and
single excited states, respectively. The symmetry irreducible
representations of all significant triplet and singlet transitions are
labeled in red and blue fronts in the spectra, respectively. All
calculated transition energies were shifted by 0.35~eV and cross
sections were convoluted by the Lorentzian function with the full width
at half-maximum of 0.05~eV. \textbf{b}, Schematic illustration of vector
potential (blue) and corresponding magnetic field (red) for infinite and
finite plasmons without shrinking wavelength. \textbf{c}, Schematic
illustration of point group changes of the whole system (including
adsorbate and plasmon) from infinite plasmon to finite plasmon. All
symmetrical operators for the finite plasmon case are depicted. The
reduction of symmetry irreducible representations from infinite plasmon
($\text{I}_{\text{h}}$) to finite plasmon ($\text{C}_{\text{3v}}$) is
also included. The symmetry irreducible representations for both allowed
triplet and singlet final states for infinite plasmon are labeled in
magenta fonts, meanwhile, the symmetry irreducible representations for
allowed triplet and singlet final states for finite plasmon are labeled
in red and blue fonts, respectively.}
\end{figure}

\begin{figure}
\caption{\textbf{Absorption under 7~\AA{} plasmon} \textbf{a},
Calculated absorption cross sections of a single \ce{C60} adsorbed on a
spaced substrate with the 6-Ring configuration under a 7~\AA{} plasmon
with confinement factor $\eta_0$ of 50 (bottom) and 100 (top). In
calculated spectra, the red and blue areas represent the contributions
from triplet and single excited states, respectively. The symmetry
irreducible representations of all significant triplet and singlet
transitions are labeled in red and blue fronts, respectively, in the
spectra. All calculated transition energies were shifted by 0.35~eV and
cross sections were convoluted by the Lorentzian function with the full
width at half-maximum of 0.05~eV. \textbf{b} Calculated absorption
images for 1$^3\text{H}_{\text{g}}$, 1$^1\text{T}_{\text{1g}}$, and
1$^1\text{H}_{\text{g}}$ (from left to right) excited states mainly
contributed by the transitions from highest occupied molecular orbital
to lowest unoccupied molecular orbital of a single \ce{C60} adsorbed on
a spaced substrate with the 6-Ring configuration under a 7~\AA{} plasmon
without shrinking wavelength. The scanning plane is around 2~\AA{} above
the \ce{C60}. The solid lines represent the skeleton of \ce{C60} and the
values are the relative maximum absorption with respect to the maximum
absorption from ${}^3\text{H}_{\text{g}}$.}
\end{figure}

\begin{methods}
\subsection{Density functional theory calculations}

A single \ce{C60} was optimized in its ground
$\text{X}^1\text{A}_{\text{g}}$ state by the \textsc{Gaussian~09} suite
of program\cite{g09d01} at the first-principles level with the
Perdew-Burke-Ernzerhof exchange-correlation functional and the Pople's
6-31+G(d) basis set. Because we considered physisorption of \ce{C60}
(Figure~1a) \textit{i.e.}, weak interaction limit between surface and
\ce{C60}, during the optimization, the symmetry of \ce{C60} was
constrained to $\text{I}_{\text{h}}$. The optimized structure is in
excellent agreement with the experimental observation in gas phase
(Figure~S3). Based on the optimized structure, adequate singlet and
triplet excited states were calculated by the time-dependent density
functional theory method at the same computational level. All calculated
vertical excitation energies were shifted by 0.35~eV to eliminate the
systematic error of the approximate
functional\cite{bauernschmitt1998jacs}.

\subsection{Absorption cross section}

The vector potential of plasmonic field was considered as a field along
$z$-axis with Gaussian distribution\cite{sun2014aom} for the
calculations of the absorption spectra. Specifically, we assume
\begin{equation}
\mathbf{A}(\mathbf{r},t)=A_0g
e^{\imath\eta_0\mathbf{k}_0\cdot\mathbf{r}}
e^{-\imath\omega{t}}\hat{\mathbf{z}}+\text{c.c.},
\label{eq:a}
\end{equation}
where $g=e^{-\alpha(\mathbf{r}-\mathbf{r}_D)^2}$ is a Gaussian function
located at $\mathbf{r}_D$ with exponent of $\alpha$ and
$\hat{\mathbf{z}}$ is the normal of the substrate surface. For plane
wave plasmonic field, we set $\alpha$ to zero. For localized plasmonic
field, the size of plasmon is determined by the full width at
half-maximum of the Gaussian function ($\Gamma$). In practical
simulations, we set $\mathbf{k}_0$ in the $xy$ plane pointing to the
observer and the angle between $\mathbf{k}_0$ and $x$-axis is set to
$45^\circ$. As a result, we have
\begin{equation}
\begin{split}
\nabla\times\mathbf{A}=&[\imath\eta_0k_y-2\alpha(y-y_D)]A_0g
e^{\imath\eta_0\mathbf{k}_0\cdot\mathbf{r}}
e^{-\imath\omega{t}}\hat{\mathbf{x}}\\
-&[\imath\eta_0k_x-2\alpha(x-x_D)]A_0g
e^{\imath\eta_0\mathbf{k}_0\cdot\mathbf{r}}
e^{-\imath\omega{t}}\hat{\mathbf{y}}+\text{c.c.},
\end{split}
\label{eq:na}
\end{equation}
where $|k_x|=|k_y|=\frac{|\mathbf{k}_0|}{\sqrt{2}}$, $\hat{\mathbf{x}}$
($\hat{\mathbf{y}}$) is the unit vector along the $x$ ($y$) direction,
and $x_D$ ($y_D$) is the $x$ ($y$) component of $\mathbf{r}_D$. Thus,
according to the Fermi's golden rule and the Wigner-Eckart
theorem\cite{mcweeny2004}, we could calculate the cross section for
singlet to singlet and singlet to triplet by
\begin{equation}
\begin{split}
\sigma_{\text{S}\to\text{S}}\propto&\frac{A_0^2}{\Delta{E}_{rg}}
\left|\langle\Psi_g|\frac{1}{2}\sum_k
\left(\hat{p}_{k,z}g_ke^{\imath\eta_0\mathbf{k}_0\cdot\mathbf{r}_k}
+g_ke^{\imath\eta_0\mathbf{k}_0\cdot\mathbf{r}_k}\hat{p}_{k}\right)
|\Psi_r^{\text{S}}\rangle\right|^2\\
\sigma_{\text{S}\to\text{T}}\propto&\frac{A_0^2g_e^2}{4\Delta{E}_{rg}}
\left(\left|\langle\Psi_g|\sum_k[\imath\eta_0k_y-2\alpha(y_k-y_D)]g_k
e^{\imath\eta_0\mathbf{k}_0\cdot\mathbf{r}_k}\hat{s}_{z,k}|
\Psi_r^{\text{T}}\rangle\right|^2\right.\\
+&\left.\left|\langle\Psi_g|\sum_k[\imath\eta_0k_x-2\alpha(x_k-x_D)]g_k
e^{\imath\eta_0\mathbf{k}_0\cdot\mathbf{r}_k}\hat{s}_{z,k}
|\Psi_r^{\text{T}}\rangle\right|^2\right),\\
\end{split}
\label{eq:sigma}
\end{equation}
where $|\Psi_g\rangle$ is the $\text{X}^1\text{A}_{\text{g}}$ ground
state, $|\Psi_r^{\text{S}}\rangle$ and $|\Psi_r^{\text{T}}\rangle$ are
the singlet and triplet excited states obtained by time-dependent
density functional theory calculations, respectively, $\Delta{E}_{rg}$
is the vertical energy between ground and excited states,
$\hat{s}_{z,k}$ is the spin operator component along the $z$-axis for
the $k$th electron, and $g_e$ is the electron $g$-factor. In practical
calculations, the $z$ component of $\mathbf{r}_D$ is around 2~\AA{}
above the topmost position of \ce{C60}. All cross sections were
calculated by the \textsc{Fasters} program\cite{fasters}.
\end{methods}


\begin{addendum}
 \item This work was supported by the National Natural Science
Foundation of China (21421063), the ``Strategic Priority Research
Program'' of the Chinese Academy of Sciences (XDB01020200), and Swedish
Research Council (VR). The Swedish National Infrastructure for Computing
(SNIC) was acknowledged for computer time.
 \item[Author contributions] Y.L. conceived the idea and supervised the
project. S.D. and Z.R. derived formulas. S.D. wrote code and
performed calculations. All authors analysed data and wrote the paper.
 \item[Additional information] Supplementary information is available in
the online version of the paper. Reprints and permissions information is
available online at www.nature.com/reprints. Correspondence and requests
for materials should be addressed to Y.L.
 \item[Competing Interests] The authors declare that they have no
competing financial interests.
 \item[Corresponding author] Correspondence to: Yi Luo (email:
yiluo@ustc.edu.cn).
\end{addendum}

\end{document}